\documentclass[twocolumn,aps]{revtex4}
\usepackage{fleqn}
\usepackage{epsf}
\usepackage{amssymb}
\def\vec#1{\ifmmode\mathchoice{\mbox{\boldmath$\displaystyle#1$}}
{\mbox{\boldmath$\textstyle#1$}}
{\mbox{\boldmath$\scriptstyle#1$}}
{\mbox{\boldmath$\scriptscriptstyle#1$}}\else
\hbox{\boldmath$\textstyle#1$}\fi}
\begin{document}
\title{Crisis--induced intermittency due to attractor--widening in a buoyancy--driven solar dynamo}
\author{Eurico Covas${}^{2}$\thanks{Web: http://www.eurico.web.com}\&
Mathieu Ossendrijver${}^{1}$\thanks{Email: mathieu@kis.uni-freiburg.de}}
\affiliation{
1. Astronomy Unit, Mathematical Sciences, Queen Mary \& Westfield
College, Mile End Road, London, United Kingdom\\
2. Kiepenheuer-Institut f\"ur Sonnenphysik, Sch\"oneckstrasse 6, D-79108 Freiburg, Germany
}
\date{\today}
\begin{abstract}
In a recent paper [M.\ A.\ J.\ H.\ Ossendrijver, A\&A {\bf 359}, 364 (2000)]
numerical simulations of a 2D mean--field model where shown to produce grand minima, typical of
the long-term behavior of solar magnetic activity. The model consisted of dynamo that features an
$\alpha$ effect based on the buoyancy instability of magnetic fluxtubes, which gives rise to
the switching back and forth from grand minima to ``regular'' solar behavior.
In this Letter, we report evidence from a time-series analysis of the model for the presence of 
crisis--induced intermittency due to attractor--widening. We support this finding
by showing that the average duration of the minima, $\langle\tau\rangle$, follows the theoretically
predicted scaling $\langle\tau\rangle \sim (C_{\delta\alpha}-C_{\delta\alpha}^*)^{-\gamma}$,
where $C_{\delta\alpha}$ is the bifurcation parameter of interest, together with other statistical
evidence.
As far as we are aware, this is the first time concrete and detailed evidence has been
produced for the occurrence of this type of crisis--induced intermittency -- due to attractor
widening -- for such dynamo models.
\end{abstract}
\maketitle
The records of past solar magnetic activity reveal an outstanding phenomenon referred to
as grand minima. During the latest of such intervals, the so-called Maunder minimum 
(1645--1715), sunspots were virtually absent. 
This and earlier grand minima 
are clearly visible in the records of cosmogenic isotopes, e.g.\ $^{14}$C and $^{10}$Be \cite{beeretal1998}). 
Timing and duration of the known grand minima are irregular.
Solar variability is also apparent e.g.\ in length and amplitude variations of the
11-year sunspot cycle. The spectral and statistical properties of solar variability are
still not well-known. Although there is some evidence for various modulations of the solar cycle \cite{gleissberg1967},
it remains unclear, due to the lack of a sufficiently
long and accurate data set, whether they are truly
periodic rather than chaotic or even purely stochastic
\cite{weiss1990}.

Intermittency has been proposed \cite{proposed} as one of the possible scenarios for 
the underlying mechanism behind the occurrence of grand
minima. Several dynamo models have been show to exhibit quasiperiodic or chaotic intermittent behavior
\cite{intermittency.in.dynamos}. In \cite{ossendrijver2000}
a case was made for a stochastically driven 2D mean--field dynamo model showing grand minima
\cite{meanfield}.
Earlier, the authors in \cite{schmittetal1996} produced grand minima
using a 1D model, in which the flux injections  were treated as an additive random source term.
The main advantage of a 2D model is that the radial structure is resolved, so that the flux
injections can be treated more realistically through the advection term.
Of course, mean-field theory cannot replace full MHD calculations \cite{theoretical_good} and is at 
best capable of capturing the most important aspects of solar dynamo action.
The purpose of the calculations was to
illustrate that grand minima are an inherent feature of a solar dynamo based on magnetic flux tubes.

In this Letter, we show concrete evidence that the switching back and forth from grand
minima to the ``regular'' 22--year cycle is a manifestation of a known dynamical process
called attractor--widening, resulting in crisis--induced intermittency
\cite{grebogietal1987}. This type of intermittency has been found in several experimental
and numerical studies \cite{wideningothers}. Other
types of crisis \cite{grebogietal1987} have also been concretely demonstrated to exist in 
dynamo models \cite{covasetal1997c,typeIcrisisPDE}. 

\section{Dynamo model}
The induction equation for the mean magnetic field $\vec{B}$ in the {\em first-order smoothing approximation}
is given by
\begin{equation}
\frac{\partial\vec{B}}{\partial t}=\nabla\times\big\{\vec{u}\times\vec{B}+\alpha\vec{B}-
\beta\,\nabla\times\vec{B}\big\}. \label{e3}
\end{equation}
All mean quantities are defined here as longitudinal averages, so that the mean field is the axisymmetric component of the actual field.
The $\alpha$ term parametrises the effect of helical motions. In the solar dynamo,
its main effect is to create a poloidal field from a toroidal field. 
The total $\alpha$ effect consists of a buoyancy-driven term in the overshoot layer,
which exists only if $|\vec{B}|$ exceeds a threshold value, and a
small-scale kinematic $\alpha$ effect in the convection zone,
which is modeled as a random forcing term.
The flow $\vec{u}$ consists of differential rotation (chosen to model
helioseismological inversion results \cite{schouetal1998}), one updraft at a fixed low
latitude, and downdrafts, with
a finite lifetime, placed at random chosen latitudes.

For a detailed description of the model, the reader is referred to 
\cite{ossendrijver2000} ('model B'). Its main features are as follows.
The dynamo equation (\ref{e3}) is solved in a spherical shell between $r=0.5R_\odot$ and $r=R_\odot$, consisting
of a layer with convective overshooting between $r_1=0.6R_\odot$ and $r_2=0.7R_\odot$, and a convection zone above
it.
The magnetic field strength is expressed in units of $B_0$, the threshold value for the buoyancy instability.
Time is measured in terms of $R^2/\beta_0$, the magnetic diffusion time for the convection zone (see below).

\section{Results}

The numerical integration is carried out using an implicit 2D code developed
by D. Schmitt and T. Prautzsch of G\"ottingen. The grid size is set to 61 points in the radial direction,
and 51 in the latitudinal direction.

\begin{figure}[!htb]
\centerline{\epsfxsize=8.10cm \epsfbox{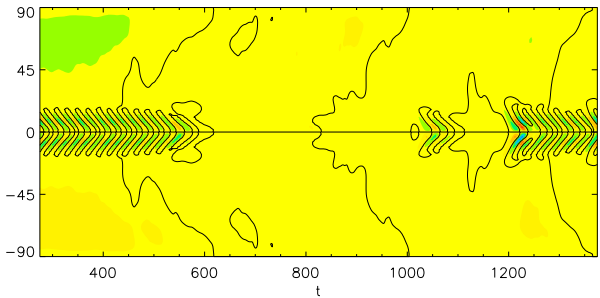}}
\caption{\label{butterfly}
Butterfly diagram of the toroidal field at $r=0.625R_{\odot}$
for parameters $C_{\omega}=2\cdot 10^4$, $C_{\alpha 1}=-0.05$, $C_{\alpha 2}=-1$,
$C_{\delta\!\alpha}=100$, $C_{\rm d}=-0.1$, and $C_{\rm b}=0.014$ (from Ossendrijver
{\protect \cite{ossendrijver2000}} -- model B). $C_\omega$ measures the
strength of the differential rotation,  $C_{\alpha 1}$ the
strength of the $\alpha$ effect on the overshoot layer,
$C_{\delta\!\alpha}        $               the strength of the mean $\alpha$ effect
in the convection zone,  $C_{\alpha 2}$ the strength of the fluctuating $\alpha$
effect in the convection zone, $C_d$ the strength of the downdrafts and $C_b$
             the strength of the updrafts. Time is in years.
                                    }
                              \end{figure}

Fig.\ \ref{butterfly} shows a butterfly diagram for the mean toroidal magnetic field
of a solution with grand minima (model B, Fig.\ 4 in \cite{ossendrijver2000}).
The numerical solution features intervals of ordinary cyclic dynamo action interrupted 
by intervals of irregular duration without cycles, reminiscent of the grand minima of 
solar activity. The ordinary solar cycle is characterized by belts of magnetic activity
on both hemispheres, which have opposite polarity, and migrate towards the equator during 
the course of a cycle. At the solar minimum, the old belt disappears, and a new belt of 
opposite polarity is formed at higher latitudes. Note that the present schematic model is 
not designed to reproduce all known detailed features of the solar cycle but some
of the most important aspects, particularly the regular butterfly diagram and the existence
of grand minima.

\begin{figure}[!htb]
\centerline{\epsfxsize=8.50cm \epsfbox{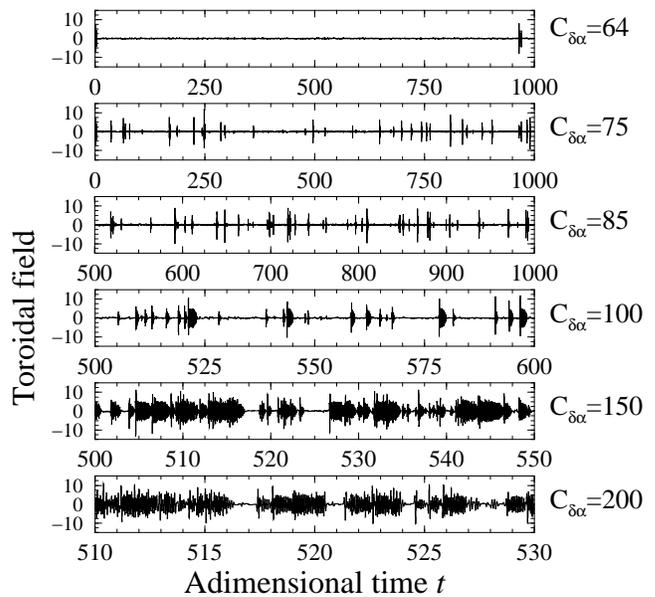}}
\caption{\label{time.series}
Times series of the toroidal field, in units of $B_0$,
at $r=0.625 R_{\odot}$ and $\theta=0.47\pi$. Except the bifurcation parameter $C_{\delta\!\alpha}$,
all parameters are as in Fig.\ {\protect \ref{butterfly}}. As
$C_{\delta\!\alpha}$ is increased beyond the critical value,
$C_{\delta\!\alpha}^*\sim 59$, the frequency of the shutdowns (grand minima) decreases rapidly.
}
\end{figure}

Fig.\
\ref{time.series} shows the evolution of the solutions as one parameter of interest,
$C_{\delta\alpha}$, is changed. The shutdown periods correpond to grand minima
and the active period correpond to the (noisy) solar cycle. As can be easily seen, if $C_{\delta\alpha}$ is raised, the
bursting becomes more and more frequent. The critical parameter value $C_{\delta\alpha}^*$
for which the smaller attractor, corresponding to the minima, starts to burst was found to be around
$C_{\delta\!\alpha}^*\sim 59$. This is indicative of the existence of some form of crisis, which
we substantiate below. 
According to the theory \cite{grebogietal1987}, around the critical parameter value
$C_{\delta\!\alpha}^*\sim 59$ the attractor
collides in phase space with the stable manifold of some unstable periodic orbit.
The excursions or burst phases seen in Fig.\ \ref{time.series} result in the widening of the
attractor after this crisis.

\begin{figure}[!htb]
\centerline{\epsfxsize=8.10cm \epsfbox{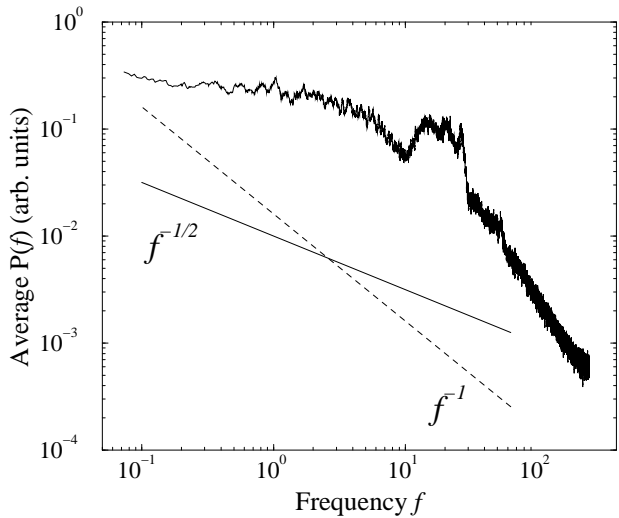}}
\caption{\label{power.spectra}
Power spectra of the time series in Fig.\ {\protect \ref{time.series}} with
$C_{\delta\!\alpha}^*\sim 100$.
Note the comparison with known scalings of power spectra for solutions of systems showing
Type I intermittency or on-off intermittency.
It shows that the time series cannot, in principle, be due to these types of
intermittency.
}
\end{figure}

\begin{figure}[!htb]
\centerline{\epsfxsize=8.10cm \epsfbox{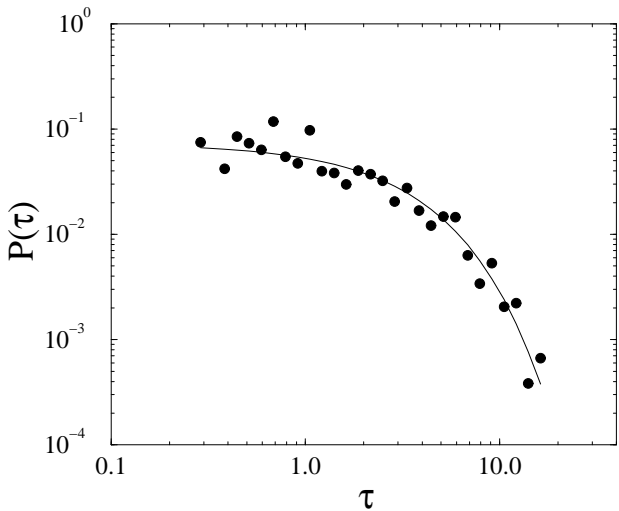}}
\caption{\label{exponential.scaling}
Scaling of the probability distribution of the duration of shutdown phases for
the time series in Fig.\ {\protect \ref{time.series}}, with
$C_{\delta\!\alpha}^*\sim 100$. Good agreement is found
for both the toroidal and poloidal (not shown) numerical obtained distribution with the known
theoretical scaling,
$\tau\sim e^{-\tau/\langle\tau\rangle}$.
}
\end{figure}

Phenomenologically, the time series could be interpreted in terms of 
various different types of intermittency. In particular, the time series could be mislead as solutions of
systems exhibiting on--off intermittency \cite{plattetal1993}, or Type I/II/III Pomeau--Manneville intermittency
\cite{pomeauetal1980}. 
However, we have eliminated this possibility by analysing both the power
spectra (see Fig.\ \ref{power.spectra}) and the probability distribution of the shutdown phases
(see Fig.\ \ref{exponential.scaling}).

It has been shown \cite{1overf} that the corresponding
power spectra for Type I intermitency have a broad-band feature whose shape obeys approximately the
inverse-power law, $P(f)\sim f^{-1}$, for $f>f_s$, where $f_s$ is the saturation frequency.
Our power spectra do not seem to show this property.
Neither do they conform to the theoretical expectation that the power spectra
for an on--off process follows a law $P(f)\sim f^{-1/2}$ \cite{venkataramanietal1996a} over
an intermediate range of frequencies. Furthermore the probability distribution of shutdown phases
(Fig.\ \ref{exponential.scaling})
does not follow the scaling $P(n)\sim n^{-\frac{3}{2}}$, as predicted for on--off intermittency \cite{plattetal1993}
but the theoretically predicted scaling \cite{grebogietal1987}
\begin{equation}
\tau\sim e^{-\tau/\langle\tau\rangle}.
\end{equation}

\begin{figure}[!htb]
\centerline{\epsfxsize=8.10cm \epsfbox{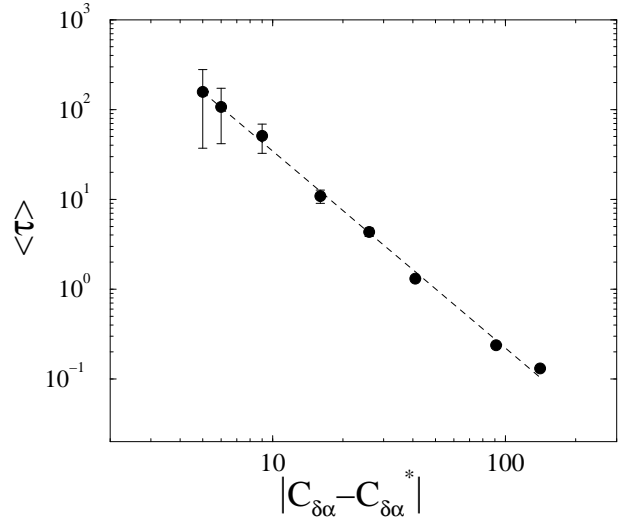}}
\caption{\label{scaling.shutdown.versus.parameter}
Scaling of the average times of shutdown (minima)
$\langle\tau\rangle$ as a function of $(C_{\delta\alpha}-C_{\delta\alpha}^*)$
for crisis--induced intermittency due to
attractor widening for model (\ref{e3}).
The slope is found to be $\gamma=2.19 \pm 0.04$ and the error bars are $\pm
\langle\tau\rangle/\sqrt{n}$, where $n$ is the number of minima found for each parameter value.
}
\end{figure}

Finally we confirm the agreement with the theoretical predictions 
for ``crisis--induced intermittency'' 
by calculating
the average time $\langle\tau\rangle$ between the bursts which obeys the scaling law \cite{grebogietal1987}
\begin{equation}
\langle\tau\rangle \sim \left|C_{\delta\alpha}-C_{\delta\alpha}^*\right|^{-\gamma},
\end{equation}
which we confirmed for our model, as can be seen in Fig.\ \ref{scaling.shutdown.versus.parameter}.
The value of $\gamma$ is found to be $\gamma=2.19 \pm 0.04$. This further reinforces the conclusion
that the type of intermittency observed here is neither Type I (for which $\gamma=-\frac{1}{2}$),
Type II or III (for which $\gamma=-1$) or on--off intermittency (for which $\gamma=-1$).

The $\gamma$ coefficient can be calculated also analytically,
as shown in \cite{grebogietal1987}. The method involves
calculating the stable and unstable manifolds of the unstable orbit
mediating the crisis. However, because our attractor seems to be of a high dimension,
the method of using the time series to reconstruct the attractor
and analyse Poincar\'e sections of such embeddings was not
useful in identifying the mediating unstable orbit. The value of $\gamma$
seems to confirm that our attractor is indeed of a high dimension, given the 
theoretically expected range $(D-1)/2\le\gamma\le(D+1)/2$ \cite{grebogietal1987}. 
This would imply $D=4$.

\section{Discussion}

We have obtained concrete evidence, in terms of time series signatures,
power spectra and dynamical scalings, to demonstrate concretely the presence of crisis--induced intermittency due
to attractor widening in a mean--field PDE
dynamo model. Despite the presence of simplifications in these models, this is
of potential importance since it shows the occurrence of another type of
intermittency (in addition to Type I Pomeau--Manneville, attractor merging intermittency
\cite{typeIcrisisPDE}
and in--out intermittency \cite{inout} recently discovered)
in these models. This may be taken as an possible indication that
more than one type of intermittency may occur in solar and stellar dynamos \cite{multiple}. Given that
the underlying model showed several important features of solar activity, such as minima of
activity and a realistic butterfly diagram, this suggest that intermittency, and in particular
the form showed here, may be the underlying nonlinear dynamical process responsible for
the observed solar magnetic activity.
\vspace{0.1cm}

EC was partially supported by a PPARC postdoctoral fellowship; MO acknowledges funding by the Deutsche
Forschungsgemeinschaft.
EC would like to thank Reza Tavakol for very useful discussions and insights.

\end{document}